\documentclass[aps,prc,showpacs,twocolumn,amssymb,superscriptaddress,nobibnotes,nofootinbib,floatfix]{revtex4}
\usepackage{graphicx}           

\begin{document}

\title{Examination of the directed flow puzzle in heavy-ion collisions}

\author{V. P. Konchakovski}
\affiliation{Institute for Theoretical Physics, University of Giessen,
  35392 Giessen, Germany}

\author{W. Cassing}
\affiliation{Institute for Theoretical Physics, University of Giessen,
  35392 Giessen, Germany}

\author{Yu. B. Ivanov}
\affiliation{Kurchatov Institute, 123182 Moscow, Russia}
\affiliation{National Research Nuclear University "MEPhI", 115409 Moscow, Russia}

\author{V. D. Toneev}
\affiliation{Joint Institute for Nuclear Research,
  141980 Dubna, Russia}

\begin{abstract}
Recent STAR data for the directed flow of protons, antiprotons, and
charged pions obtained within the beam energy scan program are
analyzed within the parton-hadron-string-dynamics (PHSD and HSD)
transport models and a 3-fluid hydrodynamics (3FD) approach. Both
versions of the kinetic approach, HSD and PHSD, are used to clarify
the role of partonic degrees of freedom. The PHSD results,
simulating a partonic phase and its coexistence with a hadronic one,
are roughly consistent with data. The hydrodynamic results are
obtained for two equations of state (EoS), a pure hadronic EoS and
an EoS with a crossover type transition. The latter case is favored
by the STAR experimental data. Special attention is paid to the
description of antiproton directed flow based on the balance of
$p\bar{p}$ annihilation and the inverse processes for $p\bar{p}$
pair creation from multimeson interactions. Generally, the
semiqualitative agreement between the measured data and the model
results supports the idea of a crossover type of quark-hadron
transition that softens the nuclear EoS but shows no indication of
a first-order phase transition.
\end{abstract}


\maketitle

\section{Introduction}

As has been widely recognized, the study of the particle azimuthal
distribution in momentum space with respect to the reaction plane is
an important tool to probe the hot, dense matter created in heavy-ion
collisions~\cite{VPS10,So10}. The directed flow refers to a
collective sidewards deflection of particles and is characterized by
the first-order harmonic $v_1$ of the Fourier expansion of the
particle azimuthal angular distribution with respect to the reaction
plane~\cite{PV98}. The second harmonic coefficient $v_2$, called
elliptic flow, and the triangular flow $v_3$ have been extensively
studied both theoretically and experimentally in the last years by
about 5 orders of magnitude in the collision energy
$\sqrt{s_{NN}}$~\cite{STAR10}. In contrast, apart from first
measurements in the early nineties and till recent times, the directed
flow was studied mainly theoretically although some experimental
information from the GSI Schwerionen Synchrotron (SIS) to CERN
Super Proton Synchrotron (SPS) energies is available~\cite{CBMbook}.

It is generally believed that the directed transverse flow is
generated early in the heavy-ion collision before a thermalization
of the degrees of freedom occurs. In particular, in the
fragmentation region (i.e.\ at large rapidity or pseudorapidity), the
directed flow is  generated during the nuclear passage
time~\cite{So97,HWW99}. The directed transverse flow therefore
probes the onset of bulk collective dynamics during thermalization,
thus providing valuable information on the pre-equilibrium
stage~\cite{SH92,KKP95,E877,NA44}. In earlier times (at moderate
beam energies) the first flow harmonic defined as
\begin{equation}
  v_1(y)=
  \left< \cos(\phi-\phi_{RP})\right>=
  \left<v_x/\sqrt{v_x^2+v_y^2}\right>
\end{equation}
with respect to the reaction plane $\phi_{RP}$ was characterized
differently, i.e., by the mean transverse momentum per particle
projected on the reaction ($x-z$) plane $\left<p_x(y)/N\right>$ in the
center-of-mass system which differs from the $v_1$ harmonic
component. Unfortunately, it is not possible to convert or directly
compare $v_1$ data to the earlier $p_x/N$ analysis. The NA49
Collaboration~\cite{NA49} has measured the flow coefficient $v_1$ for
pions and protons at SPS energies and a negative $v_1(y)$ slope was
observed by the standard event plane method for pions. Often, just the
slope of $v_1(y)$ at midrapidity has been used to quantify the
strength of the directed flow.

At BNL Alternating Gradient Synchrontron (AGS) energies $E_{lab}\lesssim$
11.5 A~GeV, the $v_1$ dependence has a characteristic S-shape
attributed to the standard $\left<p_x(y)/N\right>$ distribution. The
projected average momentum $\left<p_x(y)\right>$ grows linearly with
rising rapidity $y$ between the target and projectile fragmentation
regions. Conventionally, this type of flow -- with positive derivative
$dv_1/dy$ -- is called normal flow, in contrast to the antiflow for
which $dv_1/dy <$ 0~\cite{E877-7,NA49,RR97,HWW99}. At these moderate
energies the slope of $v_1(y)$ at midrapidity $F$ is observed to be
positive for protons and significantly smaller in magnitude and
negative for pions~\cite{E877-7,NA49,WA98}. The smooth fall off of
this function with beam energy is reasonably reproduced by the
available hadronic kinetic models (see the comparison in
Ref.~\cite{INN00}).

The shape of the  rapidity dependence $v_1(y)$ is of special
interest because the directed flow at midrapidity may be modified
by the collective expansion and reveal a signature of a phase
transition from normal nuclear matter to a quark-gluon plasma
(QGP). This is commonly studied by measuring the central rapidity
region that reflects important  features of the system evolution
from its initial state. The predicted $v_1(y)$ flow coefficient is
small close to midrapidity with almost no dependence on
pseudorapidity. However, as first demonstrated in
Refs.~\cite{Ri95, Ri96}, the 3D hydrodynamic expansion   with an
equation of state (EoS) including a possible phase transition
exhibits some irregularity in the evolution of the system. When
including a first order phase transition  this leads to a local
minimum in the proton excitation function of the transverse
directed flow at  $E_{lab}\approx$ 8 A~GeV. Such a first-order 
transition leads to a softening of the EoS and
consequently to a time-delayed expansion. The existence of this
``softest point'' of  the EoS at a minimum of the energy density
$\varepsilon_{SP}$ leads to a long lifetime of the mixed phase and
consequently in a prolonged expansion of matter~\cite{HS94}.
Presently, the critical energy density (or latent heat for a 
first-order transition at finite quark chemical potential) is not well
known and estimates vary from 0.5 GeV/fm$^3$ to 1.5
GeV/fm$^3$~\cite{HS94,ST95,MO95,RG96,RPM96}. A softest point at
$\varepsilon_{SP}\sim$ 1.5 GeV/fm$^3$ should give a minimum in the
directed flow excitation function at $E_{lab}\sim$ 30
A~GeV~\cite{HS94,ST95}. In the case of ideal hydrodynamics the
directed proton flow $p_x$ shows even a negative $v_1$ (``$v_1$
collapse'') between $E_{lab}=$ 8 A~GeV and 20 A~GeV~\cite{St05} and
with rising energy increases back to a positive flow. The ideal
hydro calculations suggest that this ``softest point collapse'' is
at $E_{lab}\sim$ 8 A~GeV but this was not confirmed by
available AGS data~\cite{St05}. However, a linear extrapolation of
the AGS data indicates that a collapse of the directed proton flow
might be at $E_{lab}\approx $ 30 A~GeV. However, this minimum
in the given energy range is not supported in the two-fluid model
with a phase transition~\cite{INN00}.

This finding was further developed in more detail in the AGS-SPS
energy range. It was demonstrated that at these energies the event
shape resembles an ellipsoid in coordinate space, tilted by an angle
$\Theta$ with respect to the beam axis. This ellipsoid expands
predominantly orthogonal to the bouncing-off direction given by
$\Theta$, forming a so-called ``third component''~\cite{CR99} or
``antiflow component''~\cite{Br00}. In addition to the deep minimum at
$E_{lab}\approx$ 8 A~GeV a clear maximum was observed at
$E_{lab}\approx $ 40 A~GeV~\cite{Br00} exhibiting a
characteristic ``wiggle''~\cite{SSV00} in the $v_1$ excitation
function. For high-energy nucleus-nucleus collisions, a combination of
space-momentum correlations of radial expansion together with the
correlation between the position of a nucleon in the nucleus and its
stopping, results in a very specific rapidity dependence of directed
flow: a reversal of the sign in the midrapidity region~\cite{SSV00},
in other words, the directed flow changes sign three times. A similar
rapidity dependence of the directed flow could be developed due to a
change in the matter compressibility if a QGP is
formed~\cite{CR99,Br00,BS02}. Although being in good agreement with
experimental data for many global observables, the three-fluid
hydrodynamic model~\cite{IRT06} with a purely hadronic EoS fails to
describe the directed flow at energies above $E_{lab}\sim$ 40
A~GeV~\cite{RI06}.

Thus, in hydrodynamic calculations~\cite{Br00,CR99,St05}, the
wiggle like structure in the $v_1$ excitation function appears only
under the assumption of a QGP with a first-order phase transition thus
becoming a signature of the QGP phase transition. The wiggle structure
is interpreted as a consequence of the expansion of the highly
compressed, disk-shaped system tilted with respect to the beam
direction~\cite{Br00}. A similar wiggle structure of the nucleon
$v_1(y)$ is predicted in transport models if one assumes strong but
incomplete baryon stopping together with strong space-momentum
correlations caused by transverse radial expansion~\cite{SSV00}.

While the predictions for baryon directed flow are very similar in
both hydrodynamical and transport models, the situation for the pion
directed flow is less clear. The Relativistic Quantum Molecular 
Dynamics (RQMD) model calculations~\cite{SSV00} for
Au+Au collisions at $\sqrt{s_{NN}}=$ 200 GeV indicate that shadowing
by protons causes the pions to flow dominantly with the opposite sign to
the protons, but somewhat diffused due to higher thermal velocities
for pions. Similar the Ultra-relativistic Quantum Molecular Dynamics 
(UrQMD) calculations~\cite{BS02} predict no wiggle
for pions in the central rapidity region with a negative slope at
midrapidity as observed at lower collision energies. It is argued
that directed flow, as an odd function of rapidity $y$, may exhibit a
small slope flatness at midrapidity due to a strong expansion of the
fireball being tilted away from the collision axis. If the tilted
expansion is strong enough, it can even overcome the bouncing-off
motion and result in a negative $v_1(y)$ slope at midrapidity,
potentially producing a wiggle-like structure in $v_1(y)$.

Note that although the calculations~\cite{Br00,CR99} for antiflow
and/or a third flow component are found for collisions at SPS
energies, where a first-order phase transition to a QGP might be
expected~\cite{St05}, the direct reason for the negative slope is
the strong, tilted expansion, which may also be important at top
BNL Relativistic Heavy Ion Collider
(RHIC) energies. The directed flow  at $\sqrt {s_{NN}}=$ 200 GeV with
a tilted source as the initial condition is predicted to be small
near midrapidity with very weak dependence on pseudorapidity.
Calculations involving a QGP phase with a first-order phase
transition suggest that $v_1(y)$ may exhibit a characteristic
``wiggle''~\cite{CR99,St05,Br00,SSV00,BS02}. In this case - in
contrast to the observed sideward deflection pattern at lower
energy, where the sign changes only at midrapidity - the directed
flow changes sign three times, not counting a possible sign change
near beam rapidities. In these calculations the wiggle structure is
interpreted as a consequence of the expansion of the system, which
is initially tilted with respect to the beam direction; the
expansion leads to the above-mentioned antiflow or third flow
component.

It is an experimental challenge to measure accurately  $v_1(y)$ at
RHIC energies due to the relatively small signal and a potentially
large systematic error arising from azimuthal correlations not
related to the reaction plane orientation (nonflow effects). The
first RHIC measurements of azimuthal anisotropy for charged
particles at $\sqrt{s_{NN}}=$ (62-200) GeV show that $v_1(y)$ appears
to be close to zero near midrapidity. Similar results have been
obtained by the STAR~\cite{STAR-v1}, PHOBOS~\cite{PHOBOS-v1}, and
PHENIX Collaborations using different correlation methods. The model
analysis of these data for nonidentified hadrons is in reasonable
agreement with experiment and shows no wiggle
structure~\cite{BW10,To12}. Generally, similar conclusions follow
from the analysis of the $v_1(y)$ excitation functions in a large
energy range carried out within different macroscopic (hydro with
hadronic, two-phase and chiral transition EoS
~\cite{BW10,MSM11,CHIRAL11}) and microscopic (UrQMD and multiphase
transport~\cite{BW10,UrQMD06,AMPT10})  models that definitely show
that systematic measurements with higher precision for identified
hadrons and more developed models are needed.

The interest in the directed flow $v_1(y)$  has recently been
enhanced considerably due to new STAR data obtained in the framework
of the beam energy scan (BES) program~\cite{STAR-14}. The directed
flow of identified hadrons -- protons, antiprotons, and positive and
negative pions -- has been measured first with high precision for
semicentral Au+Au collisions in the energy range
$\sqrt{s_{NN}}=$ (7.7-200) GeV. These data provide a promising basis
for studying direct-flow issues as discussed above and have been
addressed already by the Frankfurt group~\cite{SAP14} limiting
themselves to the energy $\sqrt{s_{NN}}<$ 20 GeV where hadronic
processes are expected to be dominant. However, the authors of
Ref.~\cite{SAP14} did not succeed in describing the data and in obtaining
conclusive results which led to the notion of the ``directed flow
puzzle''. Our study aims to analyze these STAR results in the whole
available energy range including in particular antiproton data. Here
we use two  complementary approaches: the kinetic transport [the
parton-hadron string dynamics (PHSD)] approach and relativistic
three-fluid hydrodynamics (3FD) with different equations of state.

We start with a short presentation of the PHSD approach and its
hadronic version HSD (without partonic degrees of freedom) and then
analyze the BES data in terms of both transport models to
explore where effects from partonic degrees of freedom show up.
Furthermore, we make comparisons also with predictions of other kinetic models
in Sec.~\ref{sec:PHSD} while in Sec.~\ref{sec:3FD} a similar analysis
is performed within a collective model, i.e., the 3FD.
Our findings are summarized in Sec.~\ref{sec:conclusions}.

\section{Directed flow in microscopic approaches}
\label{sec:PHSD}

\subsection{Reminder of PHSD}

The PHSD model is a covariant dynamical approach for strongly
interacting systems formulated on the basis of Kadanoff-Baym
equations~\cite{JCG04,CB09} or off-shell transport equations in
phase-space representation, respectively. In the Kadanoff-Baym
theory the field quanta are described in terms of dressed
propagators with complex self-energies. Whereas the real part of the
self-energies can be related to mean-field potentials of Lorentz
scalar, vector, or tensor type, the imaginary parts provide
information about the lifetime and/or reaction rates of timelike
particles~\cite{Ca09}. Once the proper complex self-energies of the
degrees of freedom are known, the time evolution of the system is
fully governed by off-shell transport equations for quarks and
hadrons (as described in Refs.~\cite{JCG04,Ca09}). The PHSD model
includes the creation of massive quarks via  hadronic string decay --
above the critical energy density $\sim$0.5 GeV/fm$^3$ -- and quark
fusion forming a hadron in the hadronization process. With some
caution, the latter process can be considered as a simulation of a
crossover transition because the underlying EoS in PHSD is a
crossover~\cite{Ca09}. At energy densities close to the critical
energy density the PHSD describes a coexistence of this quark-hadron
mixture. This approach allows for a simple and transparent
interpretation of lattice QCD results for thermodynamic quantities
as well as correlators and leads to effective strongly interacting
partonic quasiparticles with broad spectral functions. For a review
of off-shell transport theory we refer the reader to
Ref.~\cite{Ca09}; PHSD model results and their comparison with
experimental observables for heavy-ion collisions from the lower
SPS to RHIC energies can be found in
Refs.~\cite{Ca09,To12,KBC12,Linnyk2011}. In the hadronic phase, i.e.,
for energies densities below the critical energy density, the PHSD
approach is identical to the hadron-string-dynamics (HSD)
model~\cite{EC96,PhysRep,CBJ00}.

\begin{figure}[thb]
\includegraphics[width=0.48\textwidth]{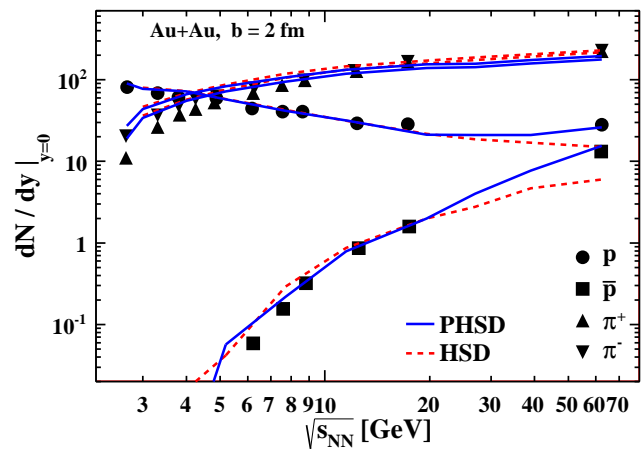}
\caption{(Color online) Particle abundance at midrapidity calculated
  for central collisions $b=$ 2 fm in the HSD (dashed lines) and PHSD
  (solid lines) models. The experimental data are from a compilation
  of Ref.~\cite{ABS06} complemented by recent data from the STAR
  Collaboration~\cite{Zhu12} and the latest update of the compilation
  of NA49 results~\cite{BGL,BM11}.}
\label{fig:en-dens}
\end{figure}

The HSD approach formally can be written as a coupled set of
transport equations for the phase-space distributions $f_h(x,p)$
of hadron $h$, which includes the real part of the scalar and
vector hadron self-energies. The hadron quasiparticle properties
here are defined via the mass-shell constraint with effective
masses and momenta. In the HSD transport calculations we include
nucleons, $\Delta$'s, $N^*(1440)$, $N^\star(1535)$, $\Lambda$, $\Sigma$,
and $\Sigma^\star$ hyperons, $\Xi$'s, and $\Omega$'s as well as
their antiparticles. High-energy inelastic hadron-hadron
collisions are described by the FRITIOF model~\cite{Fritiof},
where two incoming hadrons emerge from the reaction as two excited
color singlet states, i.e., ``strings''. The excitation functions for
various dynamical quantities as well as experimental observables
from SIS to RHIC energies within the HSD transport approach can be
found in Refs.~\cite{PhysRep,CBJ00,Brat04}.

Figure~\ref{fig:en-dens} illustrates how the hadron multiplicity
$dN/dy(y=0)$ at midrapidity is reproduced within the PHSD (solid
lines) and HSD (dashed lines) kinetic approaches. We point out that
the antiproton abundance is a crucial issue. In the AGS-SPS 
low-energy range ($\lesssim$20 GeV) both models agree quite reasonably
with experiment, including the antiproton yield. The enhancement of
the proton and antiproton yield at $\sqrt{s_{NN}}=$ 62 GeV in PHSD
relative to HSD can be traced back to a larger baryon/antibaryon
fraction in the hadronization process. At lower energies this
agreement is reached by taking into account the $p\bar p$
annihilation to three mesons (e.g., $\pi$, $\rho$, and $\omega$) as well as
the inverse channels employing detailed balance as worked out in
Ref.~\cite{Ca02}. These inverse channels are quite important; in
particular, at the top SPS energy this inverse reaction practically
compensates the loss of antiprotons due to their
annihilation~\cite{Ca02}. At lower SPS and AGS energies the
annihilation is dominant due to the lower meson abundancies;
however, the backward channels reduce the net annihilation rate. We
mention that the multiple-meson recombination channels are not
incorporated in the standard UrQMD transport model~\cite{Bass}. The
proton multiplicities are reproduced rather well in the PHSD and HSD
approaches but the multiplicity of charged pions is slightly
overestimated for $\sqrt{s_{NN}}\lesssim$ 10 GeV. This discrepancy is
observed also in other transport models~\cite{Br99,LCLM01} and is the
subject of separate investigations.

\subsection{Directed flow from microscopic dynamical models}

\begin{figure}[thb]
\includegraphics[width=0.48\textwidth]{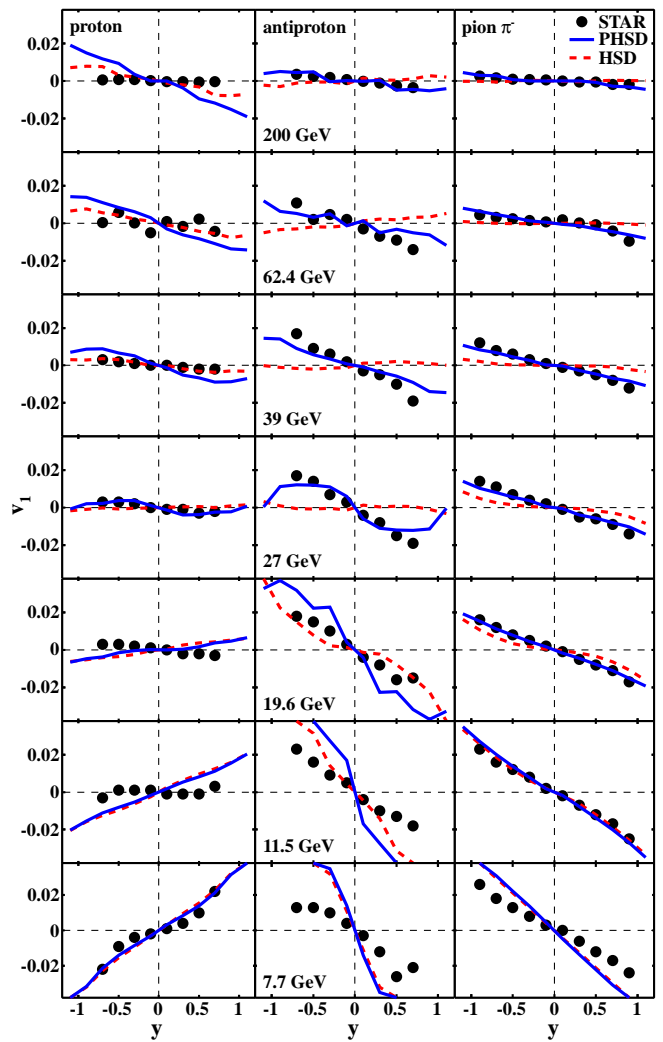}
\caption{(Color online) The directed flow $v_1(y)$ for protons,
  antiprotons as well as negative pions from 10-40\% central Au+Au
  collisions at different collision energies from $\sqrt{s_{NN}}=$ 7.7
  to 200 GeV from HSD (dashed lines) and PHSD (solid
  lines). Experimental data are from the STAR
  Сollaboration~\cite{STAR-14}.}
\label{fig:spectra}
\end{figure}

The whole set of directed flow excitation functions for protons,
antiprotons and charged pions from the PHSD and HSD models is presented 
in Fig.~\ref{fig:spectra} in comparison to the measured
data~\cite{STAR-14} including early STAR results for the two highest
energies. The initial states in the PHSD and HSD are simulated
on an event-by-event basis taking into account fluctuations in the
position of the initially colliding nucleons and fluctuations in the
reaction plane. This procedure is identical to that in the study of
the elliptic flow in Ref.~\cite{KBC12}. The average impact parameter
for the selected events is $b=$ 7 fm. In the simulations the
experimental acceptance $0.2\leq p_T \leq 2$ GeV/c is taken into
account for all hadrons~\cite{STAR-14}.

At first glance, both models -- in particular the PHSD  --
correctly reproduce the general trends in the differential $v_1(y)$
with bombarding energy: the $v_1(y)$ slope for protons is positive
at low energies ($\sqrt{s_{NN}}\le$ 20 GeV) and approaches zero with
increasing energy while antiprotons and pions have negative slopes,
respectively, in the whole energy range. In more detail: for protons
the directed flow distributions are in reasonable agreement with
the STAR measurements in the whole range of the collision energies
considered (except for $\sqrt{s_{NN}}=$ 11.5 and 200 GeV). However, $v_1(y)$
for antiprotons agrees with the data only for the highest energies
where baryon-antibaryon pairs are dominantly produced by
hadronization. This becomes evident from a comparison to the HSD
results with $v_1(y) \approx 0$. The shape of the $v_1(y)$
distribution for antiprotons starts progressively to differ from the
measured data if we proceed from $\sqrt{s_{NN}}=$ 11.5
to 7.7 GeV. In the lower energy range the HSD and PHSD results get
very close which indicates the dominance of hadronic reaction
channels (absorption and recreation). The direct flow
distributions for negative and positive pions are close to each
other and also begin to disagree with experiment in the same range
of low collision energies as for antiprotons (see
Fig.~\ref{fig:spectra}). Again the PHSD results are very close to
the experimental measurements at higher energies while the HSD
results deviate more sizeably, thus stressing the role of partonic
degrees of freedom in the entire collision dynamics. The clear
overestimation of the ${\bar p}$ and $\pi^-$ slopes at
$\sqrt{s_{NN}}=$ 7.7 GeV demonstrates that the heavy-ion dynamics
is not yet fully understood within the string/hadron picture  at
the lower energies.

\begin{figure}[thb]
\includegraphics[width=0.48\textwidth]{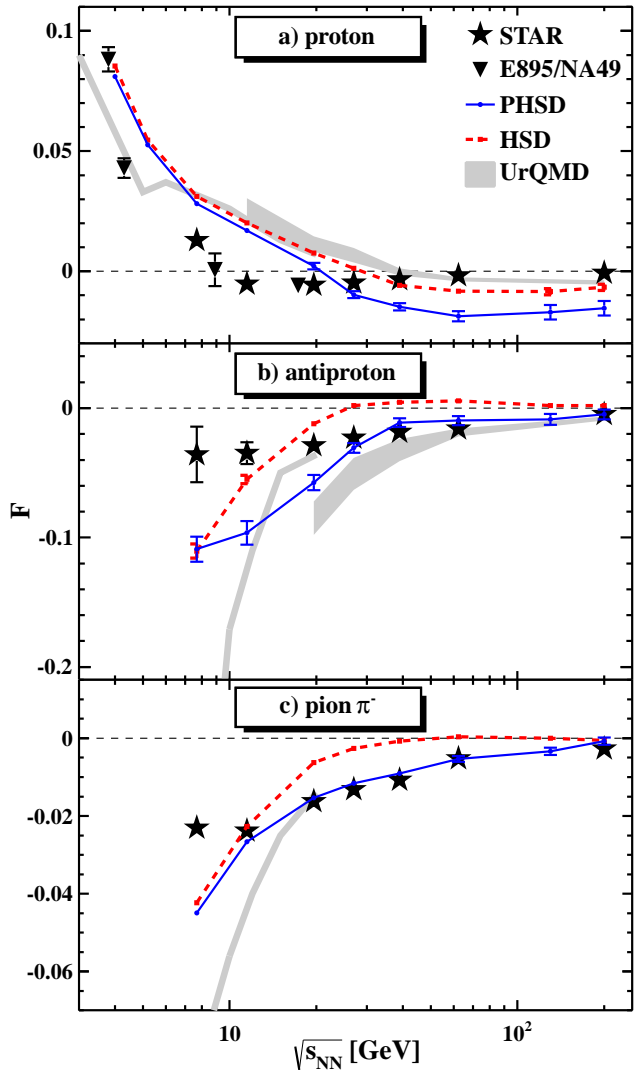}
\caption{(Color online) The beam energy dependence of the directed
  flow slope at midrapidity for protons, antiprotons and charged pions
  from semicentral Au+Au collisions. The shaded band corresponds to
  the UrQMD results as cited in Ref.~\cite{STAR-14}. The experimental data
  are from the STAR Collaboration~\cite{STAR-14} along with results of
  prior experiments using comparable cuts~\cite{NA49,E895}.}
\label{fig:slope}
\end{figure}

The characteristic slope of the $v_1(y)$ distributions at
midrapidity, $\frac{d v_1}{dy}|_{y=0}=F$, is presented in
Fig.~\ref{fig:slope} for all cases considered in
Fig.~\ref{fig:spectra}. In a first approximation the $v_1$ flow in
the  center-of-mass system may be well fitted by  a linear function
$v_1(y)=F\ y$ within the rapidity interval $-0.5<y<0.5$. A  cubic
equation is also used,
\begin{equation}
  v_1(y)=Fy+Cy^3~,
\label{cub}
\end{equation}
to obtain an estimate of the uncertainty in extracting the coefficient
$F$. The error bars in Fig.~\ref{fig:slope} just stem from the
different fitting procedures. Note that the energy axis in
Fig.~\ref{fig:slope} is extended by adding experimental results for
$\sqrt{s_{NN}}=$ 62 and 200 GeV~\cite{STAR-14}. This representation is
more delicate as compared to $v_1(y)$ in Fig.~\ref{fig:spectra}. For
protons there is a qualitative agreement of the HSD ahd PHSD results with
the experiment measurements: the slope $F>0$ at low energies, however,
exceeding the experimental values by up to a factor of about 2; the
slope crosses the line $F=$ 0 at $\sqrt{s_{NN}}\sim$ 20 GeV, which is
twice larger than the experimental crossing point, and then stays
negative and almost constant with further energy increase. However,
the absolute values of the calculated proton slopes in this high
energy range are on the level of --(0.01-0.02), while the measured
ones are about --0.005. The standard UrQMD model results, as cited in
the experimental paper~\cite{STAR-14} and in the more recent
theoretical work~\cite{SAP14}, are displayed in Fig.~\ref{fig:slope}
by the wide and narrow shaded areas, respectively. These results for
protons are close to those from the HSD model and essentially overestimate
the slope for energies below $\sim$ 30 GeV but at higher energy become
negative and relatively close to the experiment. The predictions for
the pure hadronic version of the transport model HSD [dotted lines in
Fig.~\ref{fig:slope}(a)] slightly differ from the PHSD results, which
overpredict the negative proton slope at higher RHIC energies.

\begin{figure}[thb]
\includegraphics[width=0.48\textwidth]{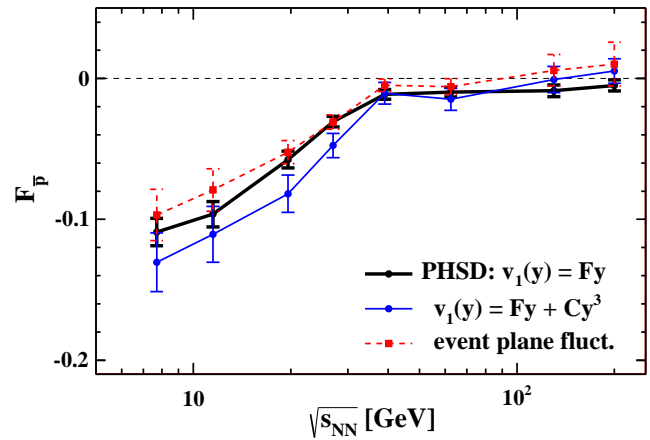}
\caption{(Color online) Excitation function of the antiproton slope
  calculated in the PHSD model with (dotted line) and without (solid line)
  including fluctuations of the reaction plane. The dotted line
  corresponds to a use of the cubic equation (\ref{cub}) for the slope
  calculation.}
\label{fig:fluct}
\end{figure}

For the antiproton slopes we again observe an almost quantitative
agreement with the BES experiment~\cite{STAR-14}: with increasing
collision energy the HSD and PHSD slopes grow and then flatten above
20-30 GeV. The HSD results saturate at $v_1(0)=0$, while the PHSD
predictions stay negative and in good agreement with experiment [see
Fig.~\ref{fig:slope}(b)]. It is noteworthy to point out that these
PHSD predictions strongly differ from the UrQMD results which no
longer describe the data for $\sqrt{s_{NN}}\lesssim$ 20 GeV but are in
agreement with the measurements for higher energies. This disagreement
might be attributed to a neglect of the inverse processes for
antiproton annihilation~\cite{Ca02} in UrQMD as described above.

\begin{figure}[thb]
\includegraphics[width=0.48\textwidth]{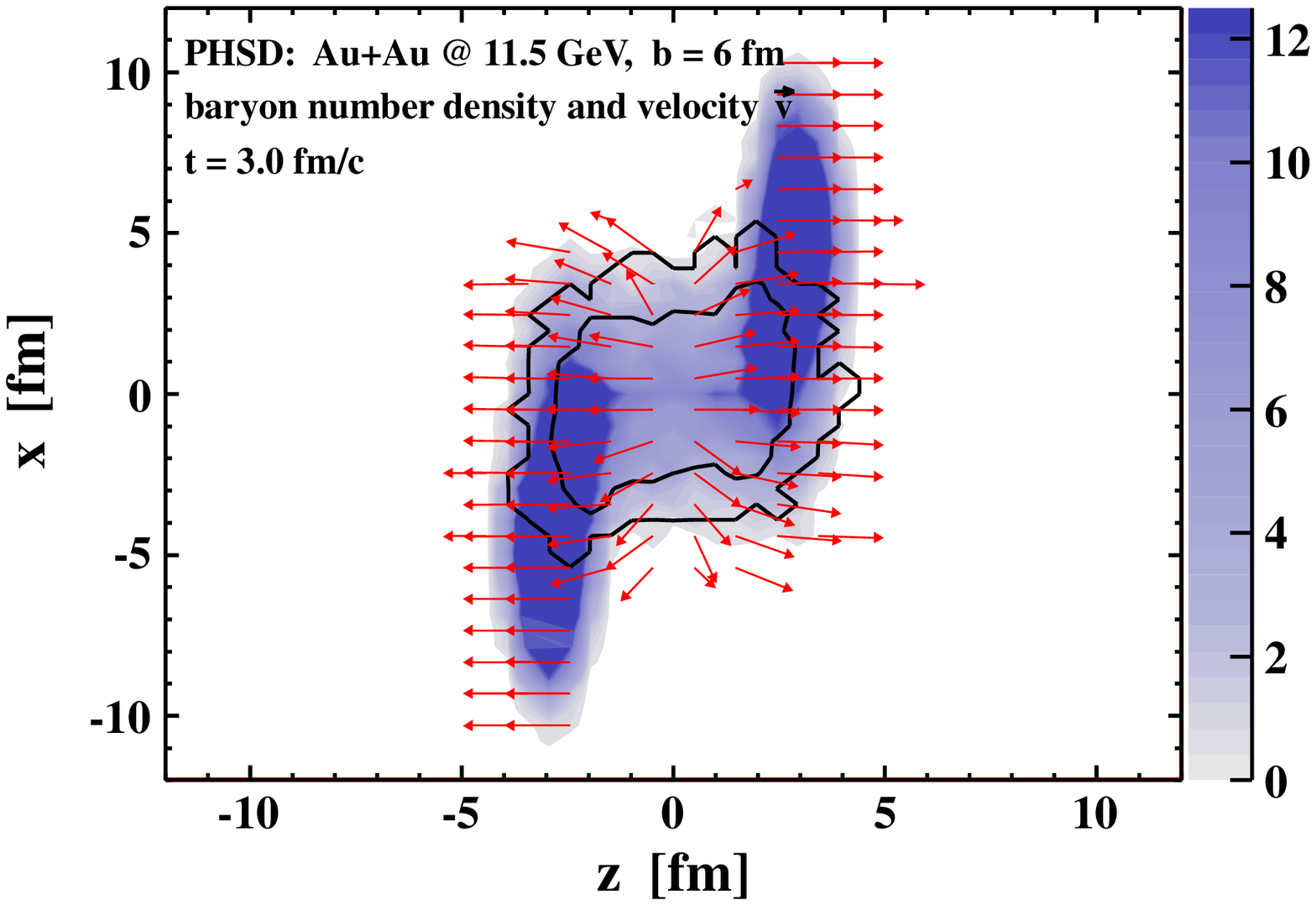}
\includegraphics[width=0.48\textwidth]{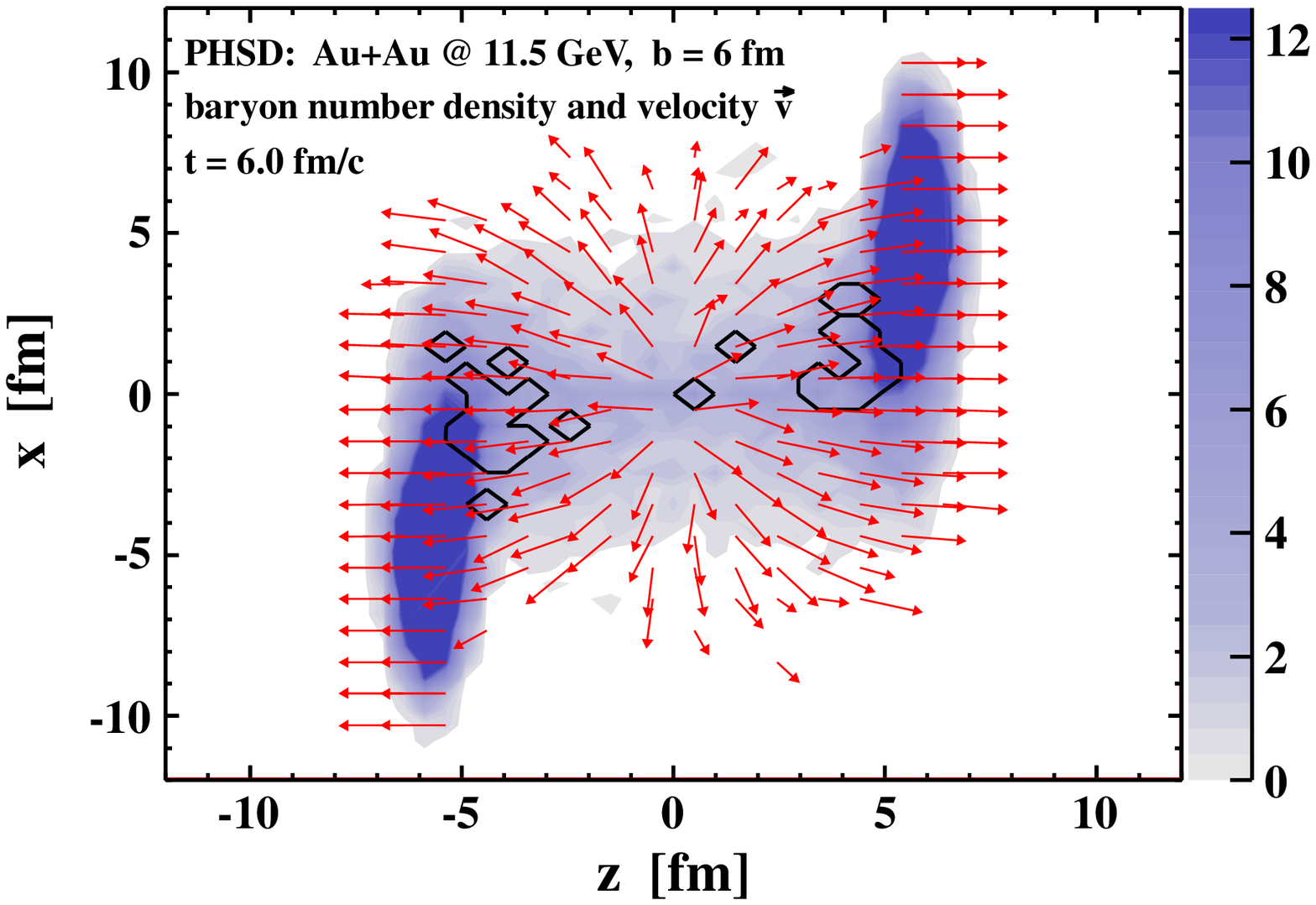}
\caption{(Color online) Snapshots of the baryon energy density
  distribution in the PHSD model at the time $t=$ 3 and 6 fm/c
  for Au+Au collisions and $\sqrt{s_{NN}}=$ 11.5 GeV. The energy
  density scale is given on the right side in GeV/fm$^3$. The solid
  curves display parton density levels for 0.6 and 0.01
  partons/fm$^3$. The arrows show the local velocity of baryonic
  matter (in relative units).}
\label{fig:dens}
\end{figure}

The differences between the calculations and experimental data become
apparent for the charged pion slopes at $\sqrt{s_{NN}}\lesssim$ 11
GeV: the negative minimum of the charged pion slope is deeper than the
measured one. The HSD and PHSD results practically coincide at low
energy (due to a minor impact of partonic degrees of freedom) but
dramatically differ from those of the UrQMD model for
$\sqrt{s_{NN}}\lesssim$ 20 GeV [see Fig.\ref{fig:slope}(c)]. This
difference might be attributed again to a neglect of the inverse
processes for antiproton annihilation in UrQMD.

As noted before, we have taken into account fluctuations of the
reaction plane which have an influence on the determination of the
$v_1$ slopes. The influence of reaction plane fluctuations on the
slope is illustrated in Fig.~\ref{fig:fluct} for the case of
antibaryons and improves the agreement with
experiment~\cite{STAR-14}. The correction due to fluctuations is not
large enough, although it acts in the right direction. We note in
passing that in the case of protons and charged pions this effect is even
smaller. Furthermore, as is seen from the same figure, the use of a
linear or cubic approximation for the fit of the $v_1(y)$
distributions around midrapidity practically does not influence the
slopes $F$ but changes the error bars.

The appearance of negative $v_1$ slopes can be explained by the
evolution of the tilted ellipsoid-like shape of the participant
zone. This situation is illustrated in Fig.~\ref{fig:dens} by PHSD
calculations and was  assumed in Refs.~\cite{CR99,Br00}. Snapshots
of the velocity profile are shown for times $t=$ 3 and 6 fm/c
for semi peripheral Au+Au (11.5 GeV) collisions in the background
of baryon density distributions where also  parton blobs can be
identified. Indeed, among the scattered particles there are many
which move perpendicularly to the stretched matter (antiflow) and
their multiplicity increases with time.

\begin{figure}[thb]
\includegraphics[width=0.48\textwidth]{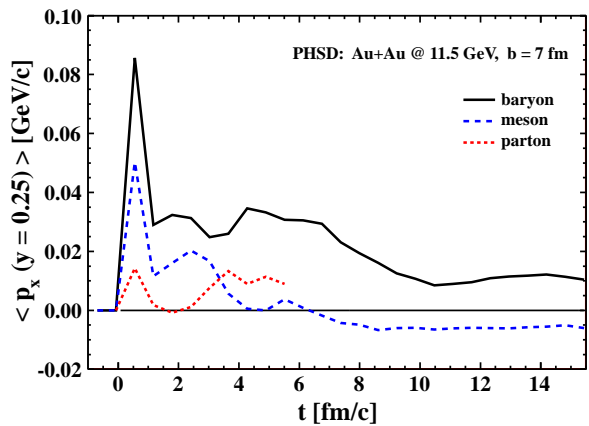}
\caption{(Color online) Evolution of the average momentum projection
  on the reaction plane for protons, pions and quarks at the shifted
  rapidity $y=0.25 \pm 0.05$. The results are given for $8.7 \times 10^4$
  PHSD events of Au+Au collisions at $\sqrt{s_{NN}}=$ 11.5 GeV.}
\label{fig:evolut}
\end{figure}

However, this component is weak and it is not clear whether these
snapshots will result in observable effects for the final slope. The
solution of this question is shown in Fig.~\ref{fig:evolut}. 
Here it is seen that the collective flow steeply rises within the
first fm/c and decreases again in time. While the flow for partons
(dotted line) stays small throughout time, the baryon flow drops to
some constant positive value and the pion flow turns negative after
$\sim 8-10$ fm/c in accordance with the results in Fig.~\ref{fig:slope}.

Thus, in agreement with the STAR experimental data, in the considered
energy range the PHSD model predicts for protons a smooth
$F(\sqrt{s_{NN}})$ function that is flattening at
$\sqrt{s_{NN}}\gtrsim$ 10 GeV and reveals no signatures of a possible
first-order phase transition as expected in
Refs.~\cite{Ri95,Ri96,St05}. For antiprotons the slope at midrapidity
manifests a wide but shallow negative minimum for
$\sqrt{s_{NN}}\approx$ 30 GeV while the measured slope is a
monotonically increasing function. It is noteworthy that the new STAR
data are consistent with the PHSD results which include a crossover
transition by default due to a matching of the EoS to lattice QCD
results.

\section{Directed flow in a macroscopic approach}
\label{sec:3FD}

\subsection{The 3FD model}

The 3FD model~\cite{IRT06} is a straightforward extension of the
two-fluid model with a radiation of direct
pions~\cite{MRS88,RIPH94,MRS91} and the (2+1)-fluid
model~\cite{Kat93,Brac97}. These models have been extended to treat
the baryon-free fluid on an equal footing with the baryon-rich ones. A
certain formation time, $\tau$, is allowed for the fireball fluid,
during which the matter of the fluid propagates without
interactions. The formation time $\tau$ is associated with the finite
time of string formation and decay and is incorporated also in the
kinetic transport models such as PHSD and HSD.

The 3FD model~\cite{IRT06} treats a nuclear collision from the very
beginning, i.e., from the stage of the incident cold nuclei to the
final freeze-out stage. Contrary to the conventional hydrodynamics,
where a local instantaneous stopping of projectile and target matter
is assumed, the specific feature of the 3FD is a finite stopping power
resulting in a counter streaming regime of leading baryon-rich
matter. The basic idea of a 3FD approximation to heavy-ion
collisions~\cite{Iv87-1,Iv87-2} is that at each space-time point a
generally nonequilibrium distribution of baryon-rich matter can be
represented as a sum of two distinct contributions initially
associated with constituent nucleons of the projectile and target
nuclei. In addition, newly produced particles, populating
predominantly the midrapidity region, are associated with a fireball
fluid. Therefore, the 3FD approximation is a minimal way to simulate
the finite stopping power at high incident energies.

\begin{figure}[thb]
\includegraphics[width=0.48\textwidth]{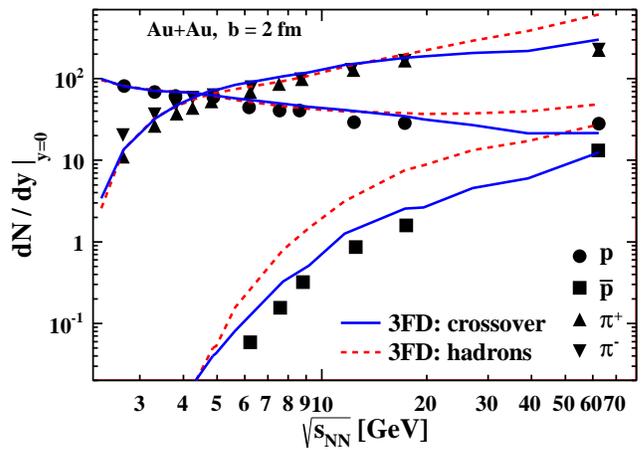}
\caption{(Color online) Particle abundance at midrapidity calculated
  for central collisions ($b=$ 2 fm) in the 3FD model with an EoS for
  a pure hadronic phase (dashed lines) and for the case of a crossover
  transition (solid lines). The experimental data are the same as in
  Fig.~\ref{fig:en-dens}.}
\label{fig:part-densH}
\end{figure}

Different EoS's can be implemented in the 3FD model in contrast to the
PHSD that incorporates only a crossover transition. In particular, in
this work we apply a purely hadronic EoS~\cite{GM79} and an EoS with a
crossover transition as constructed in Ref.~\cite{KRST06}.  In the
latter case the transition is very smooth and the hadronic fraction
(which can be treated as the order parameter) survives up to very high
densities as illustrated in Ref.~\cite{Iv13-alt1}. The physical input
of the 3FD calculations is described in detail in
Ref.~\cite{Iv13-alt1}. No tuning (or change) of 3FD-model parameters
has been done in the present study as compared to that stated in
Ref.~\cite{Iv13-alt1}.

The particle yield at midrapidity calculated within the  3FD model
is presented in Fig.~\ref{fig:part-densH}. Both the hadronic EoS
(dashed lines) and crossover EoS results (solid lines) for the
proton and pion abundancies at $\sqrt{s_{NN}}\lesssim$ 20 GeV  are in
good agreement with the experimental data and, in the case of
charged pions, in even better agreement than in the HSD and PHSD
approaches (cf.\ Fig.~\ref{fig:en-dens}). The purely hadronic EoS
definitely overestimates the antiproton yield at midrapidity in this
energy range,  while the EoS with the crossover transition quite
reasonably agrees with the experimental data. Note that the
antiprotons are mainly produced from the fireball (baryonless)
fluid~\cite{IRT06}. To a certain extent, this may be interpreted as
being due to multimeson formation of $p\bar{p}$  in equilibrium in
analogy to HSD and PHSD approaches where these channels are not in full
equilibrium. The difference between the two EoS's is clearly seen at
higher energies $\sqrt{s_{NN}}\geq$ 20 GeV, where the crossover EoS
is favorable for all hadronic species rather than only for
antibaryons ($\bar{p},\bar{\Lambda}, \bar{\Xi}^+$) as pointed out in
Ref.~\cite{Iv13-alt2}.

\subsection{Directed flow in the 3FD model}

In recent  works~\cite{Iv13-alt1,Iv13-alt2,Iv13-alt3,Iv14} an
analysis of the major part of bulk observables has been performed:
the baryon stopping~\cite{Iv13-alt1}, yields of different hadrons,
their rapidity and transverse momentum
distributions~\cite{Iv13-alt2,Iv13-alt3},
and the elliptic flow
excitation function~\cite{Iv14}. This analysis has been  carried
out for the hadronic EoS and two types of EoS with deconfinement
transition: a first-order phase transition and a crossover. It was
found that scenarios with deconfinement transitions are preferable
especially at high collision energies, though they are not perfect.

\begin{figure}[thb]
\includegraphics[width=0.48\textwidth]{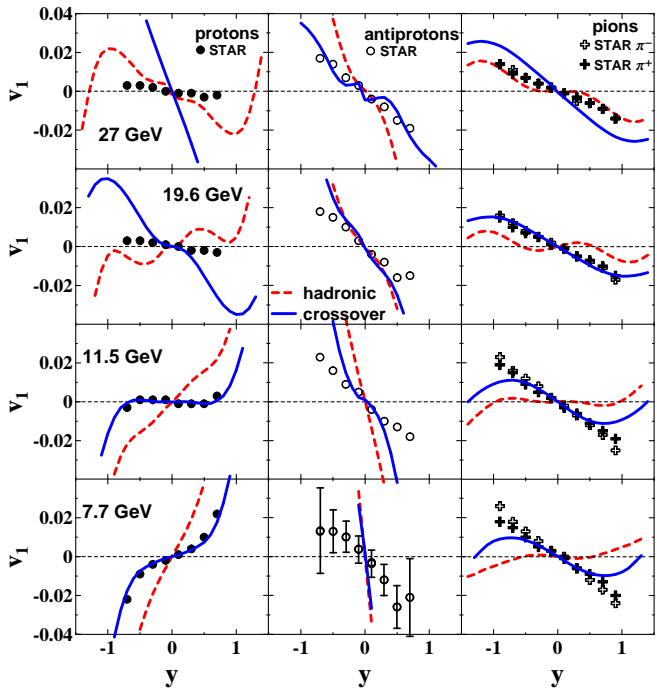}
\caption{(Color online) Rapidity distributions of the directed flow
  for protons, antiprotons, and positive and negative pions from 10 to 40\%
  central Au+Au collisions at different collision energies calculated
  within the 3FD model. The experimental data are from the STAR
  Collaboration~\cite{STAR-14}. The dashed lines correspond to a
  hadronic EoS while the solid lines stand for a crossover
  transition.}
\label{fig:spectra-3F}
\end{figure}

In this study we consider only two of the above mentioned
scenarios, i.e., the purely hadronic scenario and the crossover
one. The reason is primarily technical: It turned out that
calculations of the directed flow  are demanding and require a
high numerical accuracy. In contrast to other observables, the
directed flow is very sensitive to the step width of the
computational grid and the number of test particles.%
\footnote{A numerical ``particles-in-cell'' scheme is used in the
  present simulations; see Ref.~\cite{IRT06} and references therein
  for more details. The matter transfer due to pressure gradients,
  friction between fluids and production of the fireball fluid, is
  computed on a fixed grid (the so-called Euler step of the scheme). An
  ensemble of Lagrangian test particles is used for the calculation of
  the drift transfer of the baryonic charge, energy, and momentum (the
  so-called Lagrangian step of the scheme).}
Therefore, accurate calculations require very high memory and CPU time
and accordingly, calculations for a first-order-transition EoS have not been
completed yet. In particular, for the same reason we have failed so far
to perform calculations for energies above $\sqrt{s_{NN}}=$ 30
GeV. Note that the change of other observables, analyzed so
far~\cite{Iv13-alt1,Iv13-alt2,Iv13-alt3,Iv14}, is below 15\% as
compared to results of previous calculations.

The directed flow $v_1(y)$ as a function of rapidity $y$ at BES-RHIC
bombarding energies is presented in Fig.~\ref{fig:spectra-3F} for
pions, protons and antiprotons. As seen, the 3FD model does not
perfectly describe the $v_1(y)$ distributions. However, we can
definitely conclude that the description of the STAR data is better
with the crossover EoS than that with the purely hadronic EoS.  Note
that the negative slope at midrapidity does not necessarily assume a
QGP EoS~\cite{SSV00} once a combination of
space-momentum correlations -- characteristic of radial expansion
together with the correlation between the position of a nucleon in the
fireball and its stopping -- may result in a negative slope in the
rapidity dependence of the directed flow in high-energy nucleus-nucleus
collisions. Apparently, this is the case at $\sqrt{s_{NN}}=$ 27 GeV
with the hadronic EoS.

\begin{figure}[thb]
\includegraphics[width=0.48\textwidth]{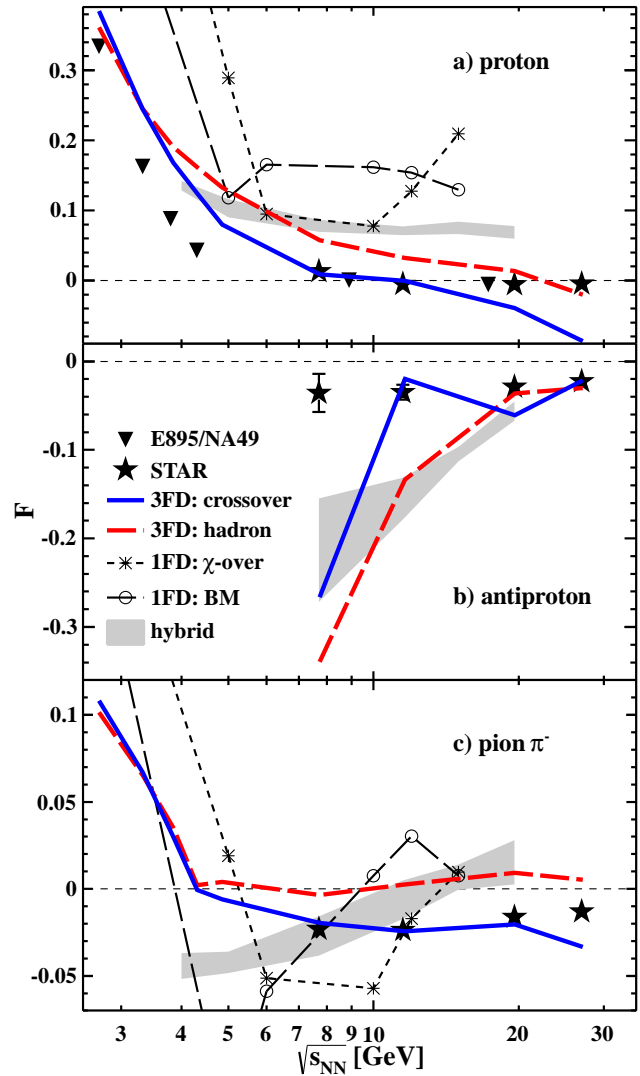}
\caption{(Color online) The beam energy dependence of the directed
  flow slope at midrapidity for protons. The lines are calculated
  within the 3FD model with a hadronic (dotted lines) and a crossover
  (solid lines) EoS. For comparison the results of calculations in
  other collective models are taken from Ref.~\cite{SAP14}. The
  experimental data are from the STAR measurements~\cite{STAR-14} and
  prior experiments with comparable acceptance
  cuts~\cite{NA49,E895,E877-00}. Note the different scales as compared
  to Fig.~\ref{fig:slope}.}
\label{fig:slope-3F}
\end{figure}

The excitation functions  for the slopes of the $v_1$ distributions
at  midrapidity are presented in Fig.~\ref{fig:slope-3F}. As noted
above, the discrepancies between experiment and the 3FD model
predictions are larger for the purely hadronic EoS (dashed line)
and, in addition, some weak substructure is observed here for
protons and pions (for example, at $\sqrt{s_{NN}}=$ 19.8 GeV). Indeed,
the agreement with the 3FD model for the crossover EoS looks better
(solid line in Fig.~\ref{fig:slope-3F}) though it is far from being
perfect. Similarly to the kinetic approaches, hydrodynamics has a
problem with the description of the low-energy behavior of the
directed flow; however, the boundary of this disagreement shifts
down to 8 GeV as compared to $\sqrt{s_{NN}}\sim$ 20 GeV in the case
of PHSD (cf.\ Fig.~\ref{fig:slope}).

In Ref.~\cite{CHIRAL11} an essential part of the STAR data (for
$\sqrt{s_{NN}}\le$ 20 GeV) is analyzed within collective
approaches: the one-fluid (1F) hydrodynamical model with a
first-order phase transition simulated by the bag model (BM) and a
crossover chiral transition ($\chi$-over), as well as within a
modern hybrid model combining  hydrodynamics with a kinetic model
in the initial and final (after-burner) stages of the collision
using both  EoS's mentioned above. The results of this work are
also displayed in Fig.~\ref{fig:slope-3F} for comparison  (the
open circles and stars).

The 3FD model predicts reasonable results for the proton slopes in the
range $\sqrt{s_{NN}}<$ 20 GeV for the crossover EoS; the pure hadronic
EoS results in a similar energy dependence but with slopes $F_p$
exceeds the experimental ones by $\sim$ 0.2. A similar behavior is
observed for the pion slope function (see Fig.~\ref{fig:slope-3F}). In
the case of antiprotons the slope for the crossover EoS (solid line in
Fig.~\ref{fig:slope-3F}) is well described above 10 GeV but it
sharply goes down with decreasing energy. For the pure hadronic EoS
the 3FD functional dependence of the antiproton slope (dashed line in
Fig.~\ref{fig:slope-3F}) looks similar but is shifted by almost 2-10
GeV towards higher energies.

The results of Ref.~\cite{SAP14} for the proton slopes in the 1FD
model overestimate the measured ones by an order of magnitude for both
chiral ($\chi$-over) and BM EoS's; appropriate results for antiprotons
are not reported. The calculational results are more definite for the
hybrid model~\cite{SAP14}: the shaded region in
Fig.~\ref{fig:slope-3F}, which covers predictions for both EoS's, is
quite close to the 3FD results with the pure hadronic EoS for protons
and antiprotons rather than to the experiment. One can conclude that
the fluid dynamical calculations presented in Ref.~\cite{CHIRAL11} are
not able to explain the observed directed flow of identified hadrons.

\subsection{Longitudinal fluctuations}

The 3FD approach describes the evolution of participants that are
defined by the initial geometry. Along with the participants there are
also spectators, i.e., nucleons that emerged from the colliding nuclei
and do not take part in any reaction with other nucleons during the
collision process and move with their initial momenta. The number of
spectators from each of the nuclei changes event-by-event and, due to
this fluctuation, the center-of-mass (cm) of the participant system
does not coincide with the collider center-of-mass system. These
event-by-event fluctuations of $y_{cm}$ are included automatically in
the kinetic approach but not in the hydrodynamic case. As noted in
Refs.~\cite{long-sc,VAC13} these fluctuations in the longitudinal cm
rapidity might be especially significant in peripheral collisions and
influence noticeably the flow characteristics.

\begin{figure}[thb]
\includegraphics[width=0.48\textwidth]{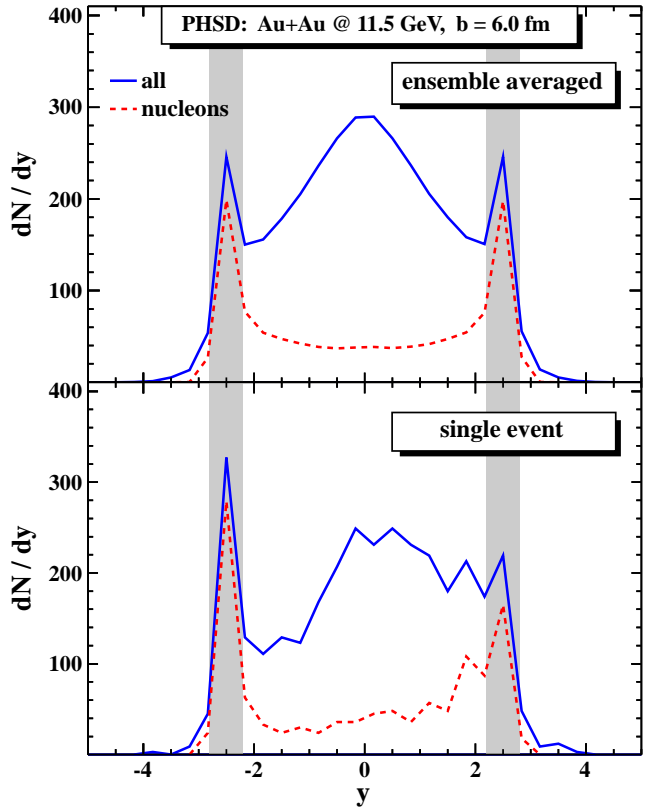}
\caption{(Color online) Rapidity distribution of nucleons (dashed
  lines) and all hadrons (solid lines) for the average over many
  events and for a single event in PHSD. 
  The shaded strips correspond to the spectator region.}
\label{fig:rap-fl}
\end{figure}

To shed some light on this issue let $P(\delta y_{cm})$ be
the probability of a fluctuation of the cm rapidity $\delta y_{cm}$
with respect to its mean value $\langle y_{cm}\rangle =0$. Then
\begin{equation}
  \langle (\delta y_{cm})^2\rangle =
  \int_{-\infty}^\infty (\delta y_{cm})^2 \ P(\delta y_{cm}) \ d\delta y_{cm}~.
\end{equation}
The $v_1$ flow at fixed $\delta y_{cm}$ is well fitted by
\begin{equation}
  v_1(y-\delta y_{cm})=F(y-\delta y_{cm})+C(y-\delta y_{cm})^3~.
\end{equation}
Then the $v_1$ flow due to fluctuations is
\begin{equation}
  v_1^{fl} = \int_{-\infty}^\infty v_1(y-\delta y_{cm})\ P(\delta y_{cm})\ d\delta y_{cm}~.
\end{equation}
Thus
\begin{equation}
  v_1^{fl} = F+3C \langle (\delta y_{cm})^2\rangle y +Cy^3~.
\label{f-slope}
\end{equation}
and the effective slope becomes $F^{fl}=F+3C \langle (\delta
y_{cm})^2\rangle$. As a rule, $F$ and $C$ are of the same order and
opposite in sign. Therefore, to significantly change $F^{fl}$
(as compared to $F$) one needs $3\langle (\delta y_{cm})^2\rangle\sim
1$, i.e.\ $\delta y_{cm}\sim$ 0.5. In Ref.~\cite{VAC13} it was estimated
that $\delta y_{cm}<$ 0.1 for midcentral Au+Au collisions, which does
not produce a noticeable effect.

\begin{figure}[thb]
\includegraphics[width=0.48\textwidth]{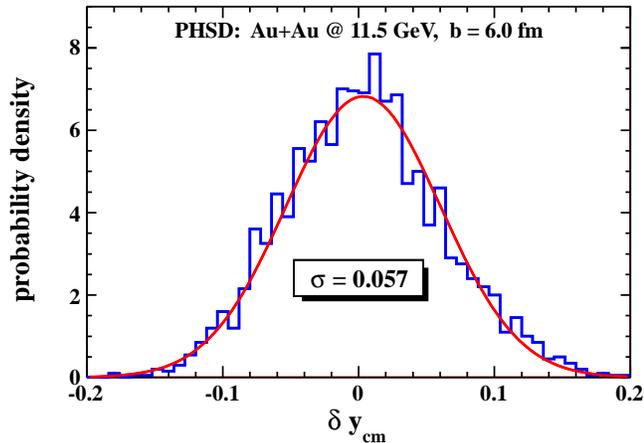}
\caption{(Color online) Particle distribution in center-of-mass
  rapidity fluctuation in PHSD. The smooth curve is the Gaussian
  approximation (\ref{gauss}).}
\label{fig:y-cm}
\end{figure}

In Fig.~\ref{fig:rap-fl} the rapidity distributions calculated for the
PHSD at energy $\sqrt{s_{NN}}=$ 11.5 GeV are presented for
the average over many events and an individual event. Note that at
this energy the PHSD~\cite{CB09} and 3FD~\cite{Iv13-alt2} predict
similar results in an approximate agreement with experiment. We
calculate the $y_{cm}$ fluctuations within the PHSD approximating the
result by a Gaussian distribution,
\begin{equation}\label{gauss}
P(\delta y_{cm})= \frac{1}{\sigma \sqrt{2\pi}} \exp(-\frac{\delta
y_{cm}^2}{2\sigma^2})~.
\end{equation}
As is seen from Fig.~\ref{fig:y-cm} the PHSD calculations of the
c.m.\ rapidity fluctuations at $\sqrt{s_{NN}}=$ 11.5 GeV give a
standard deviation $\sigma=$ 0.057 which slowly increases with energy
reaching $\sigma=$ 0.08 at 17.3 GeV. Nevertheless, influence of
fluctuations on the slope of the $v_1$ distribution remains
negligible.

\section{Conclusions}
\label{sec:conclusions}

In this study the PHSD approach has
been applied for the analysis of the recent STAR data on the
directed flow of identified hadrons~\cite{STAR-14} in the energy
range $\sqrt{s_{NN}}=$ 7.7-200 GeV. The excitation functions for the
directed flows of protons, antiprotons, and charged pions turn out to
be smooth functions in bombarding energy  without ``wiggle like''
irregularities as expected before in Refs.~\cite{
Ri95,Ri96,HS94,ST95,MO95,RG96,RPM96,St05,CR99,Br00}. Our results
differ from the standard UrQMD model at lower bombarding energies as
included in Ref.~\cite{STAR-14} and the recent theoretical analysis
in Ref.~\cite{SAP14}. The microscopic PHSD transport approach
reproduces the general trend in the differential $v_1(y)$ excitation
function  and leads to an almost quantitative agreement for protons,
antiprotons, and pions especially at higher energies. We attribute
this success to the Kadanoff-Baym dynamics incorporated in PHSD
(with more accurate spectral functions) as compared to a
Boltzmann-like on-shell transport model (UrQMD) and the account for
parton dynamics  also in this ``moderate'' energy range. The latter is
implemented in PHSD in line with an equation of state from lattice
QCD~\cite{Wuppertal}. The formation of the parton-hadron mixed phase
softens the effective EoS in PHSD and describes a crossover
transition (in line with the lattice QCD EoS). Accordingly, the PHSD
results differ from those of HSD where no partonic degrees of
freedom are incorporated. A comparison of both microscopic models
has provided detailed information on the effect of parton dynamics
on the directed flow (cf.\ Fig.~\ref{fig:spectra}).

Antiprotons have been shown to be particularly interesting. In
HSD and PHSD we include antiproton annihilation into several mesons
while taking into account also  the inverse processes of
$p\bar{p}$ creation in multimeson interactions by detailed
balance~\cite{Ca02}. Related kinetic models (including  UrQMD)
that neglect the inverse processes for antiproton annihilation at
lower energies  do not describe the data on the directed flow of
hadrons $v_1(y)$. It is noteworthy that 3FD demonstrates 
high sensitivity to the nuclear EoS and provides the best
results with a crossover for the quark-hadron phase transition
being in a reasonable agreement with the STAR results in the 
considered energy range $\sqrt{s_{NN}}<$30 GeV. Note also that
a crossover transition is implemented by default in PHSD.

Still sizable discrepancies with experimental measurements in the
directed flow characteristics are found for the microscopic kinetic
models at $\sqrt{s_{NN}}\lesssim$ 20 GeV and are common for both HSD
and PHSD (and UrQMD~\cite{Brat04}) because the partonic degrees of
freedom are subleading at these energies. We recall that the flow
observables are not only ones where the kinetic approaches have a
problem in this energy range. Another long-standing issue is the
overestimation of pion production as seen in Fig.~\ref{fig:en-dens} in
the energy regime around the ``horn'' in the $K^+/\pi^+$ meson
ratio~\cite{GG99,CBJ00}, which before has been related to a first-order
phase transition or to the onset of deconfinement~\cite{GGS11}. Our
flow analysis shows no indication of a first-order transition.
However, we have found further strong evidence
that the dynamics of heavy-ion reactions at lower SPS and AGS energies
is far from being understood especially on the hadronic level. We
speculate that extended approaches including consistently chiral
partners as well as a restoration of chiral symmetry at high baryon
density and/or temperature might lead to a solution of the problem as
well as precise experimental studies at the Facility for Antiproton
and Ion Research (FAIR) and the Nuclotron-based Ion Collider Facility
(NICA)~\cite{CBMbook}.

\begin{acknowledgments}
The authors are thankful to E. L. Bratkovskaya for illuminating
discussions and valuable suggestions. This work in part was supported
by the LOEWE Center HIC for FAIR as well as BMBF. Y.B.I. was partially
supported by Grant No.\ NS-932.2014.2.
\end{acknowledgments}


\end{document}